\documentclass[
preprintnumbers,
preprint,
a4paper,
amsmath, 
amssymb, 
floatfix,
nofootinbib,
]{revtex4}

\usepackage{graphicx}

\begin{document}

\title{Initial data for a black string and a Kaluza-Klein bubble: \\ Space-dependent compactification radius}

\author{Hirotaka Yoshino${}^{1,2}$}

\affiliation{${}^1$Department of Physics, Osaka Metropolitan University, Osaka 558-8585, Japan}

\affiliation{${}^2$Nambu Yoichiro Institute of Theoretical and Experimental Physics (NITEP),
Osaka Metropolitan University, Osaka 558-8585, Japan}

\preprint{OCU-PHYS-608, AP-GR-204}

\date{January 20, 2025}

%
%
\begin{abstract}

As the first step to explore the nonlinear dynamics of an extra dimension
in the Kaluza-Klein (KK) spacetime with black objects through numerical relativity,
we generate time-symmetric initial data of a black string and/or 
a KK bubble with space-dependent compactification radius. 
The initial data developed in this paper are classified into three types.
First, we present analytic initial data
with SO(3) symmetry whose three-dimensional section is spherically symmetric.
These initial data include a black string without a KK bubble,
a black string trapping a KK bubble, and a naked KK bubble.
Second, we present analytic initial data for multiple black strings
with varying compactification radius,
which is a natural generalization of the Brill-Lindquist initial data for four-dimensional
general relativity. 
Finally, we develop a numerical method for generating the initial data
with a black string and a KK bubble located at different positions,
which would be useful in simulating what happens 
when an expanding KK bubble meets black objects in dynamical context.
  
\end{abstract}


\maketitle

%
%

\section{Introduction}
\label{Sec:Introduction}

Various unified theories suppose the existence of extra dimensions.
As an old example, the Kaluza-Klein (KK) theory assumes the
presence of an extra dimension in order to unify general relativity and electromagnetism 
(\cite{Overduin:1997} for a review).
String theories and M-theory require the existence of six
and seven extra dimensions, respectively \cite{Becker:2006}. 
The size of extra dimensions is bounded from above by several reasons.
If the standard model fields/particles propagate extra dimensions,
the absence of excitation of the KK modes at particle colliders implies
that the size must be smaller than the Compton wavelength that corresponds
to $O(\mathrm{TeV})$ energy scale \cite{Nishiwaki:2011,Choudhury:2016}.
If the standard model fields/particles are confined on a brane,
the bound on the extra dimension size is relaxed and is determined through
the upper bound of the Planck energy \cite{Hou:2015,ATLAS:2015}, 
or the small-scale gravity experiments \cite{Murata:2014,Westphal:2020}.

Since the size of the extra dimensions is small, their effects are difficult to
be detected. Therefore, it is worth focusing attention to
phenomena such that the small extra dimensions cause strong effects
that are visible to us. One of such interesting phenomena is
the nucleation of a KK bubble that was first discovered by Witten \cite{Witten:1981}.
Consider a five-dimensional (5D) KK spacetime 
with one compactified extra dimension of topology $S^1$. 
By the double Wick rotation of the 5D Schwarzschild-Tangherlini black hole solution,
the following metric is obtained:
\begin{equation}
ds^2 \ = \  -r^2dt^2 + \frac{dr^2}{1-r_0^2/r^2}
+\left(1-\frac{r_0^2}{r^2}\right)d\chi^2 + r^2\cosh^2 t d\Omega_2^2,
\end{equation}
where the coordinate $\chi$ spans the extra dimension.
Note that the proper size of the extra dimension becomes zero at $r=r_0$. 
By requiring the absence of a conical singularity
at $r=r_0$, the $\chi$ coordinate must be periodically identified with the period of
$\Delta\chi = 2\pi r_0$. To span the connected component of the spacetime,
the coordinate range $r\ge r_0$ is sufficient: The spacetime has the ``edge''
at $r=r_0$, and this edge is not singular from higher-dimensional point of view.
The coordinate range $r<r_0$ does not span the connected component of the spacetime,
and it may be called the ``bubble of nothing'' or the ``KK bubble''. 
The two-dimensional (2D) area of the bubble surface $r=r_0$ is given by $4\pi r_0^2\cosh^2 t$,
and it expands in time for $t>0$. 
The coordinates $t$ and $r$ have the analogous properties to the
Rindler coordinates, and the Minkowski-like coordinates can be introduced.
In the Minkowski-like coordinates, the bubble region expands in time and approaches
toward the distant observers approximately at the speed of light. 
Therefore, it would be widely thought that
the formation of a KK bubble is problematic to the outside environments.

The formation of a KK bubble from the standard KK spacetime
by the classical process would require singularity formation
because the topology of the space must be changed in this process.
However, there may be a possibility that a KK bubble is nucleated
through quantum gravity process. 
Based on the estimate by Euclidean quantum gravity, 
the nucleation rate
becomes smaller as the KK compactification radius is made larger \cite{Witten:1981}.
Witten also discussed the possibility that such a process is forbidden 
even by quantum mechanically due to the presence of the fermion fields.

The properties of KK bubbles have been extensively studied.
It was pointed out that the KK bubbles can have negative total gravitational mass 
by studying the initial data \cite{Brill:1989,Brill:1991}.
The time evolution of the KK bubbles with negative gravitational mass
were performed with numerical relativity by two groups \cite{Shinkai:2000,Sarbach:2003,Sarbach:2004}
(see also \cite{Corley:1994}).
The analytic solutions of KK bubbles were systematically obtained in
\cite{Chamblin:1996}, and 
an analytic solution to the process of the collision of two KK bubbles 
was presented for lower-dimensional case, i.e., in the case of a three-dimensional (3D) spacetime
plus one compactified extra dimension \cite{Horowitz:2002}. Also, 
several interesting exact solutions to the Einstein equations 
describing KK bubbles with black objects 
were obtained for 5D KK spacetimes \cite{Emparan:2001,Elvang:2002,Iguchi:2007,Tomizawa:2007,Kunz:2008,Yazadjiev:2009-1,Yazadjiev:2009-2,Nedkova:2010,Kunz:2013}. There, it was shown that a KK bubble 
can stay in a static or stationary configuration due to gravity of black objects.
See also exact solutions \cite{Astorino:2022,Suzuki:2023,Tomizawa:2024} and initial data \cite{Copsey:2006,Copsey:2007}
for the systems of black objects with bubbles of nothing which are not
(necessarily) KK bubbles.

We now explain the interest of this paper. 
In a realistic situation, the KK bubble must drag matter as it expands.
Then, the speed of the expansion would become slower, and may even be stopped
due to gravity of matter.
In particular, it is natural to ask whether the KK bubble can continue to expand
when it meets black objects such as black strings. 
Since the previous works indicate that a KK bubble can be at stationary configuration
 when black objects are present, we may consider the possibility that the expansion
of a KK bubble would be stopped, and thus, the nucleation of a KK bubble would not be problematic.
The properties of interactions between a KK bubble and a black object
in dynamical processes must be studied by the method of numerical relativity.
The first step of such approaches is the initial data preparation.
Initial data must be prepared so that they satisfy
the initial value equations; that is, the Hamiltonian constraint
and the momentum constraints.  
The original motivation of this work was to develop
the method for generating initial data for a KK bubble and a black string
located at different positions to setup the initial configuration
of the collision process of the two objects. While we have succeeded in this study,
several other nontrivial solutions for the initial value equations 
have been obtained as by-products in the process of looking for their solutions. 
Since these initial data are also useful for studying the 
nonlinear dynamics of an extra dimension with space-dependent compactification radius, 
we would like to report them together in this paper.

The solutions to the initial value equations presented in this paper
are classified into three types.
First, analytic solutions for the initial value equations with SO(3)
symmetry will be presented. 
These solutions include a black string with space-dependent compactification radius
without a KK bubble, a KK bubble hidden by a black string horizon,
and a naked KK bubble. 
Second, we present the analytic initial data for multiple black strings with 
space-dependent compactification radius.
These initial data are a natural generalization of the Brill-Lindquist multi-black-hole
initial data in the standard four-dimensional general relativity \cite{Brill:1963}.
The property of two-black-string equal-mass initial data will be numerically
studied. Finally, we present the numerical method for
generating initial data of a KK bubble and a black string located at different positions,
and present the obtained solutions. These initial data would serve as the
initial conditions for numerical relativity simulations for the collisions of
KK bubbles and black strings.

This paper is organized as follows.
In Sec.~\ref{Sec:Initial-Value-Equations}, we present the constraint equations
for time-symmetric initial data and the metric ansatz used in this paper.
In Sec.~\ref{Sec:Spherically-symmetric}, the 
SO(3) symmetric initial data are presented,
and in Sec.~\ref{Sec:Brill-Lindquist}, the Brill-Lindquist
multi-black-string initial data are studied.
Then, in Sec.~\ref{Sec:BS-KKbubble},
the method for generating initial data of a KK bubble and a black string
located at separate positions is developed.
Section~\ref{Sec:Summary} is devoted to a summary and discussion.
Throughout this paper, the unit in which $c=G_{\rm N}=1$ is adopted,
where $c$ is the speed of light and $G_{\rm N}$ is the 
(four-dimensional) Newton gravitational constant evaluated at asymptotic region.

%
%
\section{Initial value equations}
\label{Sec:Initial-Value-Equations}

Let $\mathcal{M}(g_{ab})$ be a 5D spacetime  with a metric $g_{ab}$.
In $\mathcal{M}$, we consider a  four-dimensional (4D) initial spacelike hypersurface
$\Sigma(h_{ab}, K_{ab})$ 
with the induced metric $h_{ab}$ and the extrinsic curvature $K_{ab}$, 
where $h_{ab}$ is given by
$h_{ab} = g_{ab}+n_an_b$ with the future-directed timelike unit normal $n^a$ to $\Sigma$, 
and $K_{ab}$ is determined as
$K_{ab}= {h_a}^c\nabla_cn_b$.
The extrinsic curvature $K_{ab}$ is also written as
$K_{ab}=(1/2)\pounds_nh_{ab}$ 
where $\pounds_n$ denote the Lie derivative with respect to $n^a$.

We suppose the spacetime to be vacuum. 
The initial spacelike hypersurface
must satisfy the Hamiltonian constraint,
\begin{equation}
{}^{(4)}R+K^2-K_{ab}K^{ab} \ = \ 0,
\label{Hamiltonian}
\end{equation}
and the momentum constraint,
\begin{equation}
D_aK^a_{~b}-D_bK \ = \ 0,
\label{Momentum}
\end{equation}
where $D_a$ is the covariant derivative on $\Sigma$,
${}^{(4)}R$ is the Ricci scalar of $\Sigma$,
and $K:={K^a}_a$.
In this paper, we consider the time-symmetric initial data
(or equivalently, the momentarily static initial data),
\begin{equation}
K_{ab} = 0.
\end{equation}
Then, the momentum constraint is trivially satisfied,
and we have the equation
\begin{equation}
{}^{(4)}R \ = \ 0.
\label{Hamiltonian2}
\end{equation}

In what follows, we look for 4D spaces
that satisfy Eq.~\eqref{Hamiltonian2}. 
We will drive several kinds of initial data, and 
in each case, the 3D space is supposed
to be conformally flat, and the remaining one direction
is compactified to realize the KK spacetime.
The three-dimensional space is spanned by the
Cartesian coordinates, $(x,y,z)$, and the compactified
direction is spanned by $\chi$. We suppose the coordinate $\chi$
to be periodic with the period of $\Delta\chi$.

In Sec.~\ref{Sec:Spherically-symmetric}, we adopt the ansatz
\begin{equation}
ds^2 \ = \ \Omega(\mathbf{r})^2(dx^2+dy^2+dz^2)+\frac{F(\mathbf{r})^2}{\Omega(\mathbf{r})^2}d\chi^2,
\label{metric-ansatz1}
\end{equation}
where $\mathbf{r}$ is the abbreviation for $(x,y,z)$. 
Substituting into Eq.~\eqref{Hamiltonian2}, we obtain
\begin{equation}
\Omega\nabla^2 F + F\nabla^2\Omega \ = \ \nabla\Omega\cdot\nabla F,
\label{equation-ansatz1}
\end{equation}
where the definitions of $\nabla$ and the inner product are the same as the
ones in the standard vector analysis in a 3D flat space,
and $\nabla^2$ is the flat-space Laplacian.
This equation is symmetric with respect to exchange of $\Omega$ and $F$.
In Secs.~\ref{Sec:Brill-Lindquist} and \ref{Sec:BS-KKbubble},
we adopt the ansatz 
\begin{equation}
ds^2 \ = \ \Psi(\mathbf{r})^4(dx^2+dy^2+dz^2)+\Phi(\mathbf{r})^2d\chi^2.
\label{metric-ansatz2}
\end{equation}
In this case, the Hamiltonian constraint of Eq.~\eqref{Hamiltonian2} becomes
\begin{equation}
\Psi\nabla^2\Phi+2\nabla\Phi\cdot\nabla\Psi+4\Phi\nabla^2\Psi \ = \ 0.
\label{equation-ansatz2}
\end{equation}
In the ansatz of Eq.~\eqref{metric-ansatz2}, the compactification radius
is given by the proper length of the extra dimension divided by $2\pi$;
that is, $R(\mathbf{r})=\Delta\chi\Phi(\mathbf{r})/2\pi$, which is position-dependent.

In the KK spacetimes, the total gravitational mass is defined
by the higher-dimensional generalization of the Arnowitt-Deser-Misner (ADM) mass
for the four-dimensional spacetime. Defining the radial coordinate $r$ by $r=|\mathbf{r}|$,
the ADM mass in this system is expressed as 
\begin{equation}
M_{\rm ADM} \ = \ \frac{1}{16\pi G_5}\lim_{r\to\infty}
\oint (\partial_ih_{ij}-\partial_j h_{ii})N^jdS 
\end{equation}
in the asymptotically Cartesian coordinates $x^i$, 
where $h_{\mu\nu}:=g_{\mu\nu}-\eta_{\mu\nu}$
is the next to leading order metric, $N^i$ is the outward unit normal
to $r=\mathrm{constant}$ surface, $G_5$ is the five-dimensional gravitational constant,
and the integral is taken over
the $r=\mathrm{constant}$ surface with the topology of $S^2\times S^1$. 
In general, the structure of the extra dimension contributes to the
ADM mass \cite{Brill:1989}.
Supposing that $\Psi$ and $\Phi$ behave as 
\begin{eqnarray}
\Psi  &\approx& 1+ \frac{M/2}{r}, \\
\Phi &\approx & 1+ \frac{2\mu}{r},
\end{eqnarray} 
at the distant place
in the metric ansatz of Eq.~\eqref{metric-ansatz2}, 
the ADM mass is given by
\begin{equation}
M_{\rm ADM} \ = \ M+\mu.
\end{equation}
Note that $G_5$ is related to the Newton gravitational constant
as $G_5=G_{\mathrm N}\Delta\chi$, and we are working in the unit
where $G_{\mathrm N}=1$.

%
%
\section{SO(3) symmetric initial data}

\label{Sec:Spherically-symmetric}

In this section, we look for the solution of SO(3)-symmetric initial data 
with a space-dependent compactification radius. 
Adopting the metric ansatz of Eq.~\eqref{metric-ansatz1},
we look for solutions that
only depend on $r$. Therefore, 
the initial data have the spherical symmetry with respect to the 3D space. 
Assuming the following forms of $F$ and $\Omega$,
\begin{eqnarray}
F &=& 1+\frac{a_1}{r}+\frac{a_2}{r^2},\\
\Omega &=& 1+\frac{b_1}{r}+\frac{b_2}{r^2},
\end{eqnarray}
Eq.~\eqref{equation-ansatz1} gives only one relation,
\begin{equation}
2(a_2+b_2) \ = \ a_1b_1.
\label{relation-coefficients}
\end{equation}
Since there is one relation between four constants,
the initial data are specified by three parameters.
We restrict our attention to the case $b_1>0$ and $b_2>0$
where the positivity of $\Omega$ is guaranteed.
On the other hand, we do not require the positivity of $F$:
The existence of a region with $F<0$ means that a KK bubble
is present at the position where $F=0$ is satisfied.

In what follows, we consider the two cases separately. 
The first one is the case without a KK bubble,
and the second one is the case with a KK bubble.
These two cases are discussed in Secs.~\ref{SubSec:without-bubble}
and \ref{SubSec:with-bubble}, respectively.

\subsection{Black string initial data with varying compactification radius}
\label{SubSec:without-bubble}

In the case without the KK bubble, $F(r)$ is strictly positive throughout
$0<r<\infty$. Below, we introduce the three appropriate parameters 
that can be interpreted more intuitively rather than $a_1$, $a_2$, $b_1$, and $b_2$.
The first parameter is the mass-like quantity $M$ that is determined
from the 3D section of $\chi=\mathrm{constant}$; that is,
\begin{equation}
b_1 \ = \ M.
\label{definition-M}
\end{equation}
Note that $M$ is different from the Arnowitt-Deser-Misner (ADM) mass
because in KK spacetimes, the $O(1/r)$ part of $\chi\chi$ component of the metric
also contributes to the ADM mass.
Next, we consider the area $A_2(r)$ of a 2D sphere given by
$\chi=\mathrm{constant}$ and $r=\mathrm{constant}$.
$A_2(r)$ is given by
\begin{equation}
A_2(r) \ = \ 4\pi r^2\Omega(r)^2.
\end{equation}
This 2D area has the minimum value at 
\begin{equation}
r \ = \ r_{\rm min}:= \sqrt{b_2}.
\label{definition-rmin}
\end{equation}
We choose $r_{\rm min}$ as the second parameter.
In order to introduce the third parameter,
we consider the radius of the compactification
as the physical length of the extra dimension
divided by $2\pi$. In the asymptotic domain $r\to\infty$, 
the compactification radius $R_\infty$ is
\begin{equation}
R_\infty \ = \ \frac{\Delta\chi}{2\pi},
\end{equation}
because $F=\Omega=1$.
On the other hand, the compactification radius $R_0$ at $r=0$
becomes
\begin{equation}
R_0 \ = \ \lim_{r\to0}\frac{\Delta\chi F(r)}{2\pi \Omega(r)}
\ = \ \frac{\Delta\chi}{2\pi}\ \frac{a_2}{b_2}.
\end{equation}
Then, we have 
\begin{equation}
a_2 \ = \ \frac{R_0}{R_\infty}\ r_{\rm min}^2,
\end{equation}
and we adopt $R_0/R_\infty$ as the third parameter.
From Eq.~\eqref{relation-coefficients}, $a_1$ becomes
\begin{equation}
a_1 \ = \ \frac{2r_{\rm min}^2}{M}\left(1+\frac{R_0}{R_\infty}\right).
\end{equation}
To summarize, we have
\begin{eqnarray}
F &=& 1+2\left(1+\frac{R_0}{R_\infty}\right)\frac{r_{\rm min}^2}{Mr}
+\frac{R_0}{R_\infty}\ \frac{r_{\rm min}^2}{r^2},\label{F-expression-without-bubble}
\\
\Omega &=& 1+\frac{M}{r}+\frac{r_{\rm min}^2}{r^2}.\label{Omega-expression}
\end{eqnarray}
Note that the case $r_{\rm min}=M/2$ and $R_0/R_\infty = 1$ is the standard
Einstein-Rosen bridge with the constant compactification radius. 
If we adopt $M$ as the unit of the length, the nondimensional parameters
to specify the initial data space are $R_0/R_\infty$ and $r_{\rm min}/M$.

%
\begin{figure}
 \centering
 \includegraphics[width=0.4\textwidth]{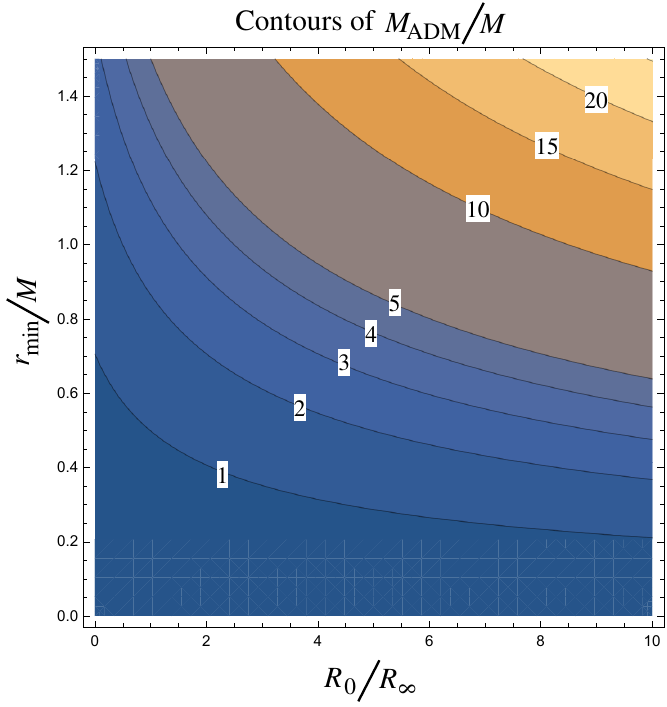}
 \caption{The contours of $M_{\rm ADM}/M$ in the case of SO(3) symmetric
 initial data without a KK bubble on the $({R_0}/{R_\infty}, r_{\rm min}/M)$-plane. 
 The value of $M_{\rm ADM}/M$
 is reduced to $1/2$ in the limit $r_{\rm min}/M\to 0$.}
 \label{Fig:without-bubble-ADM-mass}
\end{figure}
%

We discuss some of the properties of the obtained initial data.
The ADM mass of this initial space is given by
\begin{equation}
M_{\rm ADM} \ = \ \frac{M}{2}+\frac{r_{\rm min}^2}{M}
\left(1+\frac{R_0}{R_\infty}\right).
\end{equation}
The ADM mass is always greater than $M/2$, and
its value increases as $r_{\rm min}/M$ or ${R_0}/{R_\infty}$
is increased. The dependence of $M_{\rm ADM}/M$ on
the system parameters ${R_0}/{R_\infty}$ and $r_{\rm min}/M$ 
is presented in Fig.~\ref{Fig:without-bubble-ADM-mass}.

%
\begin{figure}
 \centering
 \includegraphics[width=0.4\textwidth]{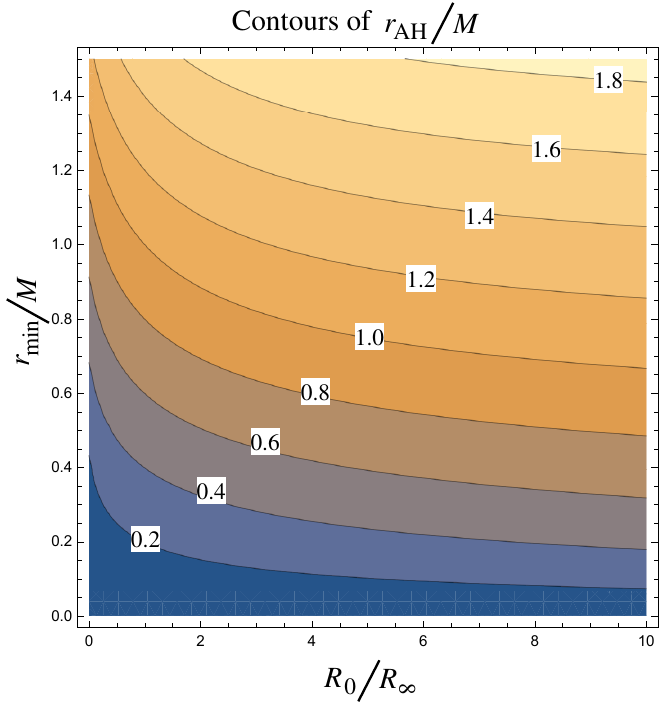}
 \includegraphics[width=0.4\textwidth]{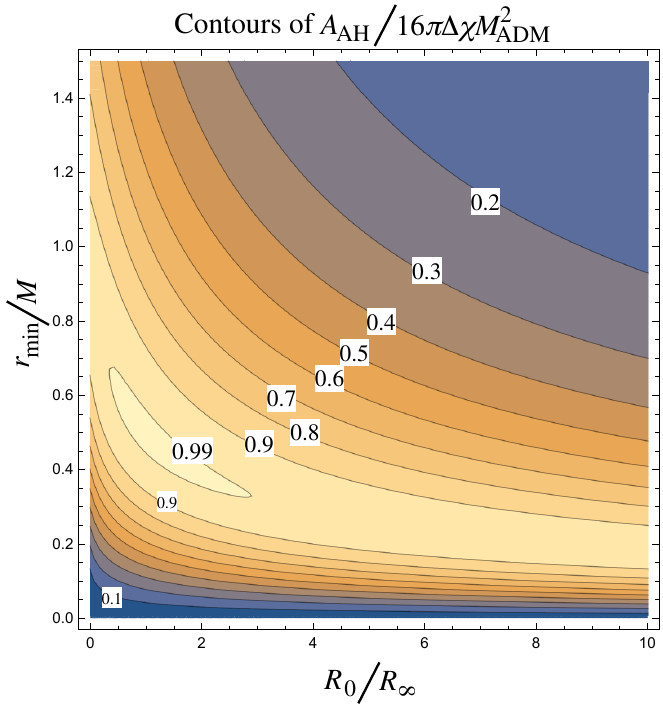}
 \caption{The property of an apparent horizon in the case of SO(3) symmetric
 initial data without a KK bubble. Left panel:
 Contours of the radial position $r_{\rm AH}/M$ on the $({R_0}/{R_\infty}, r_{\rm min}/M)$-plane.
 Right panel: Contours of the area of the apparent horizon
 normalized by that of the Schwarzschild string with the same ADM mass,
 $A_{\rm AH}/16\pi\Delta\chi M_{\rm ADM}^2$, on the $({R_0}/{R_\infty}, r_{\rm min}/M)$-plane.}
 \label{Fig:without-bubble-AH}
\end{figure}
%

Next, we look at the apparent horizon. Since the initial space
is time-symmetric, the apparent horizon coincides with
the minimal surface. The 3D area of an $r=\mathrm{constant}$ surface
is given by
\begin{equation}
A_3(r) \ = \ 4\pi \Delta\chi\ r^2F(r)\Omega(r).
\end{equation}
It is possible to determine the position of the apparent horizon,
$r=r_{\rm AH}$, 
by solving the equation $dA_3/dr \ = \ 0$. The left panel of
Fig.~\ref{Fig:without-bubble-AH} shows the contours
of $r_{\rm AH}/M$ on the $({R_0}/{R_\infty}, r_{\rm min}/M)$-plane.
The apparent horizon always exists. 
The right panel of Fig.~\ref{Fig:without-bubble-AH} shows the contours
of the area of the apparent horizon $A_{\rm AH}$ normalized
by $16\pi \Delta\chi M_{\rm ADM}^2$ on the same plane.
This normalization factor is identical to the area of the
Schwarzschild string with the same ADM mass.
Since $A_{\rm AH}/16\pi \Delta\chi M_{\rm ADM}^2$ is less than or equal to unity, 
the area of the apparent horizon is bounded from above by that of the
Schwarzschild string 
in the case without the KK bubble.

\subsection{Initial data of a naked KK bubble or a KK bubble hidden by a black string}
\label{SubSec:with-bubble}

We now turn our attention to the case with a KK bubble.
Similarly to the case without a KK bubble, we use
$M$ and $r_{\rm min}$ defined in 
Eqs.~\eqref{definition-M} and \eqref{definition-rmin}
as two of the three parameters to specify the initial data.
Instead of $R_0/R_\infty$, we use
the radial position of the bubble as the last parameter
in this case. Since $r_{\rm B}$ satisfies $F(r_{\rm B}) = 0$, we have
\begin{equation}
r_{\rm B}^2 + a_1r_{\rm B}+a_2 \ = \ 0.
\end{equation} 
Together with Eq.~\eqref{relation-coefficients}, $a_1$ and $a_2$ are solved for as
\begin{eqnarray}
a_1 &=& \frac{r_{\rm min}^2-r_{\rm B}^2}{r_{\rm B}+M/2},\\
a_2 &=& - r_{\rm B}\frac{r_{\rm min}^2+Mr_{\rm B}/2}{r_{\rm B}+M/2}.
\end{eqnarray}
As a result, we have
\begin{equation}
F \ = \ 1+\frac{r_{\rm min}^2-r_{\rm B}^2}{r_{\rm B}+M/2}\, \frac{1}{r}
- \frac{r_{\rm min}^2+Mr_{\rm B}/2}{r_{\rm B}+M/2}\, \frac{r_{\rm B}}{r^2},
\end{equation}
instead of Eq.~\eqref{F-expression-without-bubble}. 
As for $\Omega$, Eq.~\eqref{Omega-expression} is used.
We adopt $M$ as the unit of the length, and use
the nondimensional parameters $r_{\rm B}/M$ and $r_{\rm min}/M$
to specify the initial data.

%
\begin{figure}
 \centering
 \includegraphics[width=0.4\textwidth]{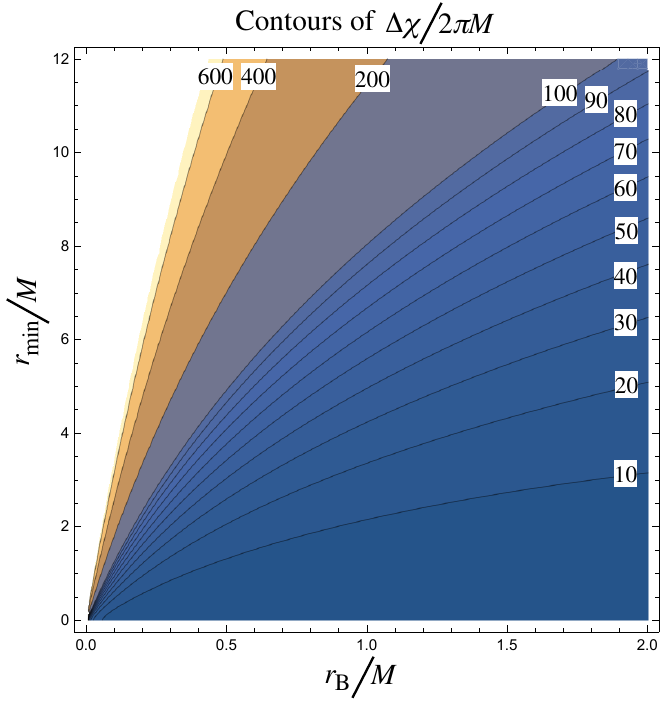} \caption{
 Contours of the period $\Delta\chi$ normalized by $2\pi M$ on the $(r_{\rm B}/M, r_{\rm min}/M)$-plane in the case with a KK bubble.}
 \label{Fig:with-bubble-period}
\end{figure}
%

In the case that the KK bubble is present,
the period $\Delta\chi$ of the $\chi$ coordinate
must be adjusted to avoid the appearance of a conical singularity
on the bubble. The metric of the 2D section of $\theta, \phi=\mathrm{constant}$
is approximately given by
\begin{equation}
ds^2 \ \approx \ \Omega^2(r_{\rm B}) 
\left[dr^2
+\frac{F^{\prime 2}(r_{\rm B})(r-r_{\rm B})^2}{\Omega^4(r_{\rm B})} d\chi^2
\right],
\end{equation}
in the neighborhood of the bubble. Therefore, the period of $\chi$ must be
\begin{equation}
\Delta\chi \ = \ \frac{2\pi\Omega^2(r_{\rm B})}{F^{\prime}(r_{\rm B})}
\ = \ \frac{2\pi}{r_{\rm B}^2}
\left(r_{\rm B}^2+Mr_{\rm B}+r_{\rm min}^2\right)\left(r_{\rm B}+\frac{M}{2}\right).
\end{equation}
Figure~\ref{Fig:with-bubble-period} shows the behavior of
$\Delta\chi/2\pi M$ on the $(r_{\rm B}/M, r_{\rm min}/M)$-plane.
$\Delta\chi/2\pi M$ is always greater than $1/2$, and becomes large as $r_B/M$ is decreased.

%
\begin{figure}
 \centering
 \includegraphics[width=0.4\textwidth]{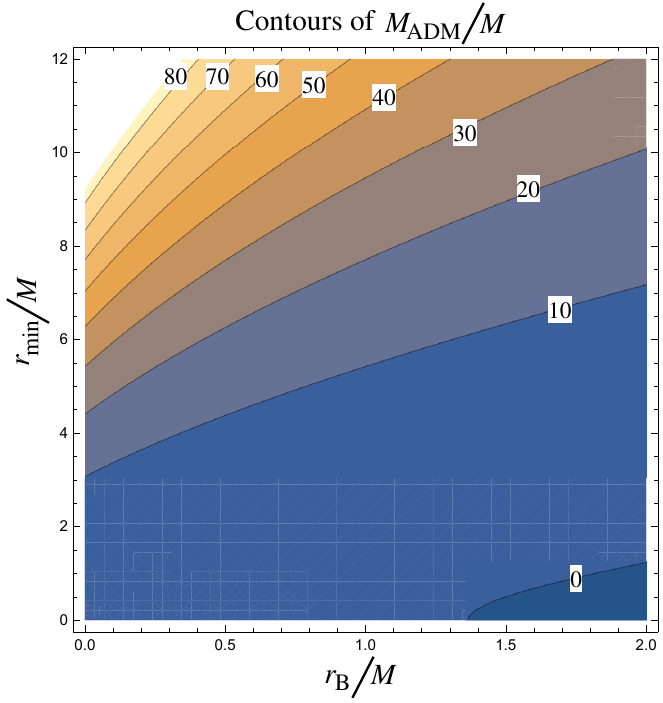} \caption{
 Contours of $M_{\rm ADM}/M$  on the $(r_{\rm B}/M, r_{\rm min}/M)$-plane in the case with a KK bubble. There appears a region of negative ADM mass.}
 \label{Fig:with-bubble-ADMmass}
\end{figure}
%

The ADM mass is given by
\begin{equation}
M_{\rm ADM} \ = \ \frac{M}{2}+\frac{r_{\rm min}^2-r_{\rm B}^2}{2r_{\rm B}+M}.
\end{equation}
It is found that the value of $M_{\rm ADM}$ becomes
negative if $r_{\rm B}$ is sufficiently large.
It is well known that the ADM mass can have negative mass
in the cases of KK bubbles \cite{Brill:1989}, and our initial data also include such configurations.
Figure~\ref{Fig:with-bubble-ADMmass} shows the contours of
$M_{\rm ADM}/M$  on the $(r_{\rm B}/M, r_{\rm min}/M)$-plane.

%
\begin{figure}
 \centering
 \includegraphics[width=0.4\textwidth]{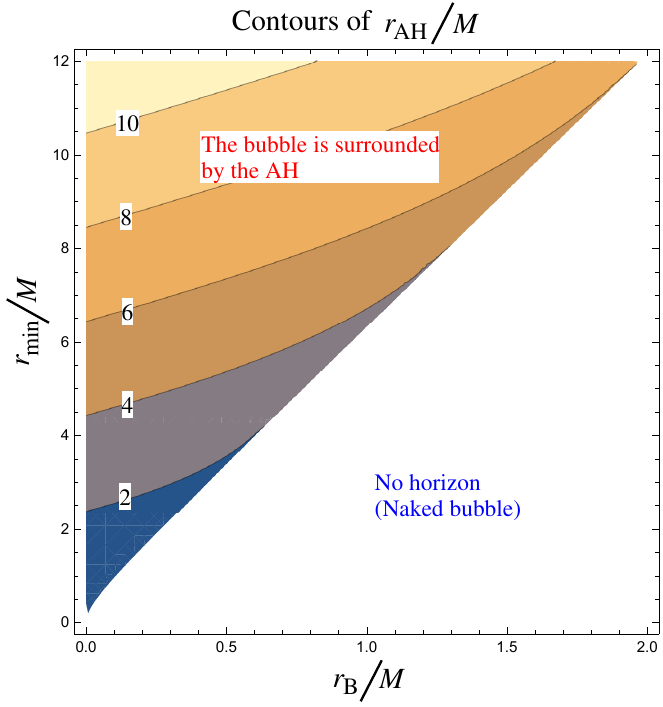} 
 \includegraphics[width=0.4\textwidth]{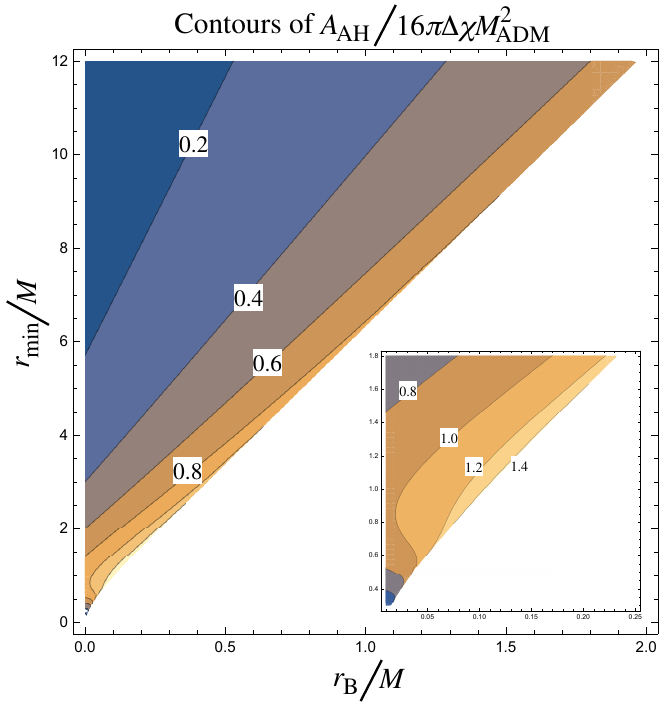} 
 \caption{The property of an apparent horizon in the case of SO(3) symmetric
 initial data with a KK bubble. Left panel:
 Contours of the radial position $r_{\rm AH}/M$ on the $(r_{\rm B}/M, r_{\rm min}/M)$-plane.
 There is a region without the apparent horizon. 
 Right panel: Contours of the area of the apparent horizon
 normalized by that of the Schwarzschild string with the same ADM mass,
 $A_{\rm AH}/16\pi\Delta\chi M_{\rm ADM}^2$, on the $(r_{\rm B}/M, r_{\rm min}/M)$-plane.
 The inset enlarges the region where $A_{\rm AH}/16\pi\Delta\chi M_{\rm ADM}^2$ exceeds unity.}
 \label{Fig:with-bubble-AH}
\end{figure}
%

The location of the apparent horizon can be found by the same method
as the case without the KK bubble. 
The left panel of Fig.~\ref{Fig:with-bubble-AH} shows the radial position $r_{\rm AH}$
of the apparent horizon on the $(r_{\rm B}/M, r_{\rm min}/M)$-plane. 
In contrast to the case without the KK bubble,
there is a region where the apparent horizon is not present; In that region,
the KK bubble is ``naked'' in the sense that it is not surrounded by the apparent horizon.
The right panel shows the contours of $A_{\rm AH}/16\pi\Delta\chi M_{\rm ADM}^2$.
There is a region where the value of $A_{\rm AH}/16\pi\Delta\chi M_{\rm ADM}^2$
exceeds unity. It is well known that $A_{\rm AH}$ becomes
larger than $16\pi\Delta\chi M_{\rm ADM}^2$ in the case of 
a black string in the Gregory-Laflamme instability, and as the result
of time evolution, a naked singularity is formed \cite{Lehner:2010,Figueras:2022}.
We may be able to expect that the similar thing could happen
also in our system, although the explicit study by the time evolution
is required for confirmation.

%
%
\section{Brill-Lindquist multi-black-string initial data}
\label{Sec:Brill-Lindquist}

In this section, we discuss the method for generating initial data
that do not possess SO(3) symmetry: The multi-black-string
initial data with varying size of an extra dimension but without 
a KK bubble.

\subsection{Solving the initial value equation}

We adopt the metric ansatz of Eq.~\eqref{metric-ansatz2}. 
The equation for $\Psi$ and $\Phi$ is given by
Eq.~\eqref{equation-ansatz2}, and it is rewritten as
\begin{equation}
\nabla\cdot(\Psi^2\nabla\Phi) + 4\Phi\Psi\nabla^2\Psi \ = \ 0.
\end{equation}
Assuming that there is the following relation between
$\Phi$ and $\Psi$,
\begin{equation}
\Phi \ = \ C+\frac{D}{\Psi},
\end{equation}
the equation is rewritten as
\begin{equation}
\left(4C\Psi+3D\right)\, \nabla^2\Psi \ = \ 0.
\end{equation}
Then, any solution to the Laplace equation $\nabla^2\Psi \ = \ 0$
gives the initial data satisfying the initial value equations.
We can easily
construct the $N$-black-string initial data by choosing the solution,
\begin{equation}
\Psi \ = \ 1+\sum_{i=1}^{N}\frac{M_i/2}{|\mathbf{r}-\mathbf{r}_{i}|},
\end{equation}
where $\mathbf{r}_i$ is the position of the $i$-th puncture (i.e., the coordinate point
that corresponds to the asymptotically flat region beyond the Einstein-Rosen bridge),
and $M_i$ is the mass-like parameter of the $i$-th black string.
This situation is very analogous to the Brill-Lindquist initial data
for the standard 4D general relativity \cite{Brill:1963} (which are sometimes called
the puncture initial data).

It is necessary to discuss the meaning of the constants $C$ and $D$.
In this section, we do not consider the case where the negative region
of $\Phi$ appears. If $\Phi$ can become negative, the KK bubble appears
at the surface where $\Phi=0$ is satisfied, but it is difficult to
avoid the appearance of a conical singularity in general.
For this reason, we require the positivity of $\Phi$.
Without loss of generality, we can require
$\Phi \to 1$ in the asymptotic limit $|\mathbf{r}|\to\infty$.
This is identical to $D=1-C$. 
In the asymptotic region, the radius of the compactification
is $R_\infty \ = \ {\Delta\chi}/{2\pi}$.
On the other hand, at the punctures, the radius of the compactification is
\begin{equation}
R_{\rm p} \ = \ \frac{C\Delta\chi}{2\pi}.
\end{equation}
This means $C=R_{\rm p}/R_\infty$. As a result, $\Phi$
is rewritten as
\begin{equation}
\Phi \ = \ \frac{R_{\rm p}}{R_\infty}
+\left(1-\frac{R_{\rm p}}{R_\infty}\right)\frac{1}{\Psi}.
\end{equation}
The ADM mass of the initial data is given by
\begin{equation}
M_{\rm ADM} \ = \ \frac{M}{4}\left(3+\frac{R_{\rm p}}{R_\infty}\right)
\end{equation} 
with $M=\sum_{i=1}^N M_i$.
In the case $R_{\rm p}/R_\infty = 1$, the initial data is the 
standard Brill-Lindquist initial data with a uniform extra dimension.
Therefore, the initial data obtained here is a natural generalization
of the Brill-Lindquist initial data to KK spacetimes,
and we call them the Brill-Lindquist black string initial data.

\subsection{Numerical study of the two-black-string initial data}
\label{Sec:BL-2BS}

Here, we study the property of the Brill-Lindquist initial data
with two equal-mass black strings focusing attention to the
apparent horizon that encloses both black strings (the common apparent horizon).
The two punctures are located at $z=\pm z_0$ on the $z$-axis,
and hence,
\begin{equation}
\Psi \ = \ 1+\frac{M/4}{\sqrt{x^2+y^2+(z-z_0)^2}}+\frac{M/4}{\sqrt{x^2+y^2+(z+z_0)^2}}.
\end{equation}
Introducing the spherical-polar coordinates $(r,\theta,\phi)$ in the standard manner,
the function $\Psi$ can be expressed as a function of $r$ and $\theta$ because
the system is axisymmetric. 
The system is also symmetric with respect to the reflection about the
equatorial plane, $\theta=\pi/2$.

Since the initial data is time symmetric, the apparent horizon is equivalent to
the minimal surface. The equation for the apparent horizon 
is
\begin{equation}
D_as^a \ = \ 0,
\end{equation}
where $s^a$ is the outward unit normal to the apparent horizon. 
Supposing that the apparent horizon is given by
$r\,=\,h(\theta)$, $s^a$ can be written as
\begin{equation}
s^a \ = \ \frac{1}{\Psi^2\sqrt{1+h^{\prime 2}/r^2}}
\left(1, \ -\frac{h^{\prime}}{r^2},\ 0,\ 0\right)
\end{equation}
in the $(r,\theta,\phi,\chi)$ coordinates.
Calculating its divergence and setting $r=h$, we find the equation for the
apparent horizon as
\begin{multline}
h^{\prime\prime} 
- \left(\frac{4\Psi_{,h}}{\Psi}+\frac{\Phi_{,h}}{\Phi}+\frac{2}{h}\right)h^2
-\left(\frac{4\Psi_{,h}}{\Psi}+\frac{\Phi_{,h}}{\Phi}+\frac{3}{h}\right)h^{\prime 2}
\\
+\left(\frac{4\Psi_{,\theta}}{\Psi}+\frac{\Phi_{,\theta}}{\Phi}+\cot\theta\right)h^\prime\left(1+\frac{h^{\prime 2}}{h^2}\right) 
\ = \ 0.
\end{multline}
If $\Phi=\mathrm{constant}$, the equation is equivalent to that for the
standard Brill-Lindquist initial data.
We have solved this equation by imposing the boundary condition
$h^\prime(0) = h^\prime(\pi/2) = 0$ using the shooting method.
The 3D area of the apparent horizon is calculated by
\begin{equation}
A_{\rm AH} \ = \ 4\pi \Delta\chi\int_0^{\pi/2}  \Psi^4\Phi h\sqrt{h^2+h^{\prime 2}}\sin\theta d\theta.
\end{equation}

%
\begin{figure}
 \centering
 \includegraphics[width=0.4\textwidth]{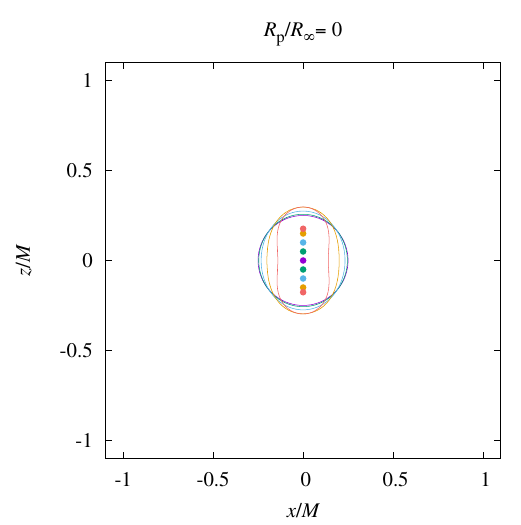} 
 \includegraphics[width=0.4\textwidth]{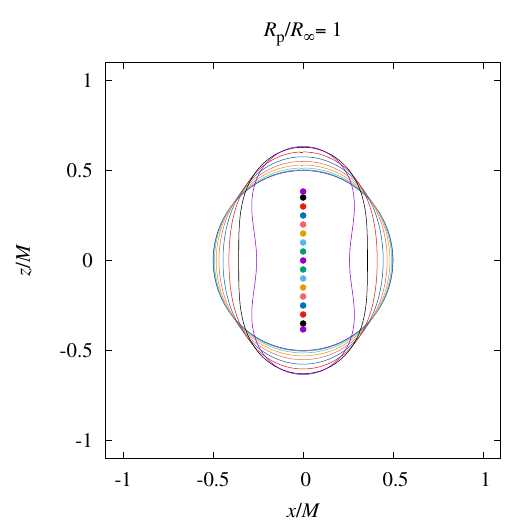} 
 \includegraphics[width=0.4\textwidth]{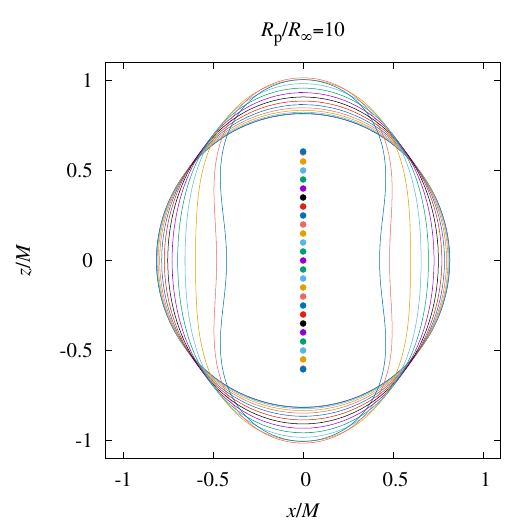} 
 \caption{Coordinate shape of the common apparent horizon
 that encloses two black strings for $R_{\rm p}/R_\infty = 0$ (top left),
 $1$ (top right), and $10$ (bottom) in the $(x,z)$-plane.
 Dots indicate the location of the punctures and the apparent horizon becomes distorted
 as $z_0/M$ is increased. Here, the cases $z_0/M=0$, $0.05$,
 $0.10$,..., and the case $z_0/M=z_0^{\rm (c)}/M$ are plotted.}
 \label{Fig:CommonAH-BrillLindquist}
\end{figure}
%

%
\begin{figure}
 \centering
 \includegraphics[width=0.6\textwidth]{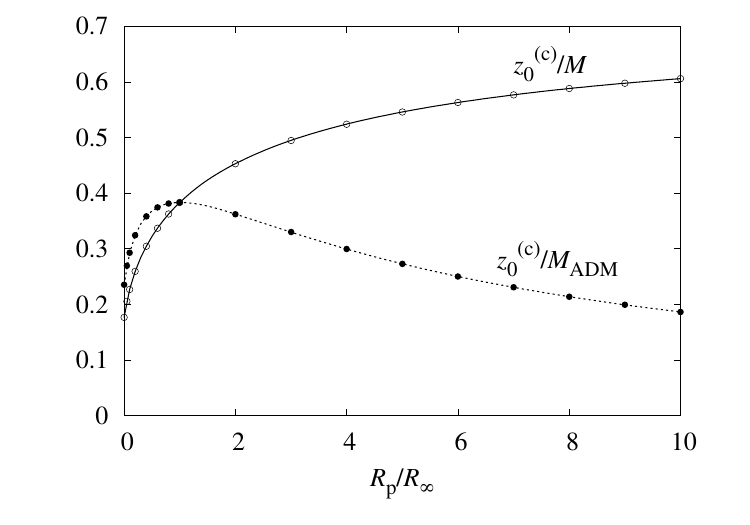} 
 \caption{The values of $z_0^{\rm (c)}/M$ (circles, $\circ$) 
 and $z_0^{\rm (c)}/M_{\rm ADM}$ (dots, $\bullet$)
 as functions of $R_{\rm p}/R_\infty$.}
 \label{Fig:CommonAH-zcrit}
\end{figure}
%

%
\begin{figure}
 \centering
 \includegraphics[width=0.6\textwidth]{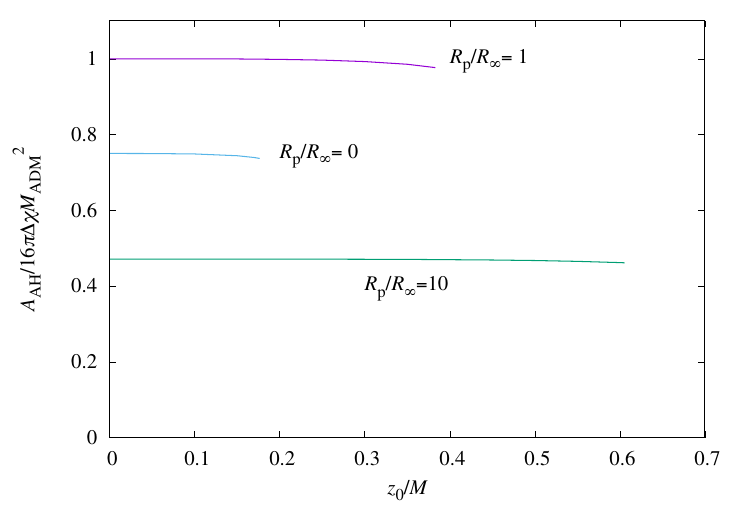} 
 \caption{The values of $A_{\rm AH}/16\pi \Delta\chi M_{\rm ADM}^2$
 as functions of $z_0/M$ for $R_{\rm p}/R_\infty=0$, $1$, and $10$.}
 \label{Fig:CommonAH-area}
\end{figure}
%

Figure~\ref{Fig:CommonAH-BrillLindquist} shows the shape of the
common apparent horizon for $R_{\rm p}/R_\infty = 0$ (left panel),
 $1$ (middle panel), and $10$ (right panel) in the $(x,z)$-plane.
Dots indicate the locations of the punctures.
In all cases, the common apparent horizon becomes distorted
as $z_0$ is increased, and there is a critical value $z_0^{\rm (c)}$ such that
no common apparent horizon can be found for $z_0>z_0^{\rm (c)}$.
The values of $z_0^{\rm (c)}/M$ are 
$\approx 0.1765$ for $R_{\rm p}/R_\infty = 0$, 
$\approx 0.3830$ for $R_{\rm p}/R_\infty = 1$,
and $\approx 0.6057$ for $R_{\rm p}/R_\infty = 10$.
There is a subtle issue in interpreting this result. 
If we regard $M$ as the characteristic scale, on the one hand, 
the coordinate shape of the apparent horizon becomes smaller 
as the value of $R_{\rm p}/R_\infty$ is decreased as indicated in Fig.~\ref{Fig:CommonAH-BrillLindquist}.
In fact, in the limit $z_0\to 0$, 
the coordinate radius of the apparent horizon is given by
\begin{equation}
\frac{r_{\rm AH}}{M} \ = \  \frac{1}{8}
\left(1-\hat{R}_{\rm p}+\sqrt{1+14\hat{R}_{\rm p}+\hat{R}_{\rm p}^2}\right),
\end{equation}
where we have introduced $\hat{R}_{\rm p}:=R_{\rm p}/R_\infty$
(i.e., the compactification radius at the pucture normalized by that at infinity)
in order to simplify the expression. 
This is a monotonically increasing function of $\hat{R}_{\rm p}$; 
The value of $r_{\rm AH}/M$ is $1/4$ for $\hat{R}_{\rm p}=0$,
$1/2$ for $\hat{R}_{\rm p}=1$, and $1$ for $\hat{R}_{\rm p}\to \infty$.
Correspondingly, the condition for the apparent horizon formation becomes stricter
as $R_{\rm p}/R_\infty$ is decreased
in the sense that $z_0^{\rm (c)}/M$ becomes smaller as shown in Fig.~\ref{Fig:CommonAH-zcrit}.
On the other hand, if we adopt the higher-dimensional point of view,
$M_{\rm ADM}$ would give a characteristic scale. In that case, we have
\begin{equation}
\frac{r_{\rm AH}}{M_{\rm ADM}} 
\ = \  
\frac{1-\hat{R}_{\rm p}+\sqrt{1+14\hat{R}_{\rm p}+\hat{R}_{\rm p}^2}}{2\left(3+\hat{R}_{\rm p}\right)}
\end{equation}
in the limit $z_0\to 0$, which takes the maximum value at $\hat{R}_{\rm p}=1$, 
and hence, we may interpret that the apparent horizon
is most easily formed when $\hat{R}_{\rm p}=1$.
This property also can be read from Fig.~\ref{Fig:CommonAH-zcrit}
in which $z_0^{\rm (c)}/M_{\rm ADM}$ is also plotted as a function of $R_{\rm p}/R_\infty$.
In any case, the condition for the apparent horizon formation
is strongly affected by the structure of the extra dimension.

Figure~\ref{Fig:CommonAH-area}
shows the 3D area $A_{\rm AH}$ of the apparent horizon
normalized by that of the Schwarzschild string with the same ADM mass,
$A_{\rm AH}/16\pi \Delta\chi M_{\rm ADM}^2$.
The value of this quantity is always less than or equal to unity
and weakly depends on $z_0/M$. 
For a fixed $R_{\rm p}/R_\infty$, the value of $A_{\rm AH}/16\pi \Delta\chi M_{\rm ADM}^2$
is a decreasing function of $z_0/M$.
In the limit $z_0/M\to 0$, this quantity is expressed as
\begin{equation}
\frac{A_{\rm AH}}{16\pi \Delta\chi M_{\rm ADM}^2} \ = \
\frac{
\left(1+14\hat{R}_{\rm p}+\hat{R}_{\rm p}^2\right)^{3/2}-\left(1+\hat{R}_{\rm p}\right)
\left(1-34\hat{R}_{\rm p}+\hat{R}_{\rm p}^2\right)}{8\hat{R}_{\rm p}\left(3+\hat{R}_{\rm p}\right)^2},
\end{equation}
whose value is $3/4$ for ${\hat{R}_{\rm p}}=0$,
$1$ for ${\hat{R}_{\rm p}}=1$ (which is a maximum value), and $0$
for ${\hat{R}_{\rm p}}\to \infty$.

%
%
\section{Initial data for a black string and a KK bubble}
\label{Sec:BS-KKbubble}

In this section, we study the numerical method for generating initial data
with both a black string and a KK bubble.
In contrast to the SO(3) symmetric case in Sec.~\ref{SubSec:with-bubble},
we consider the situation where a KK bubble is not trapped 
inside the apparent horizon of a black string:
For the initial data in this section, the black string and the KK bubble
are expected to collide after the time evolution.

\subsection{Formulation}

%
\begin{figure}
 \centering
 \includegraphics[width=0.6\textwidth]{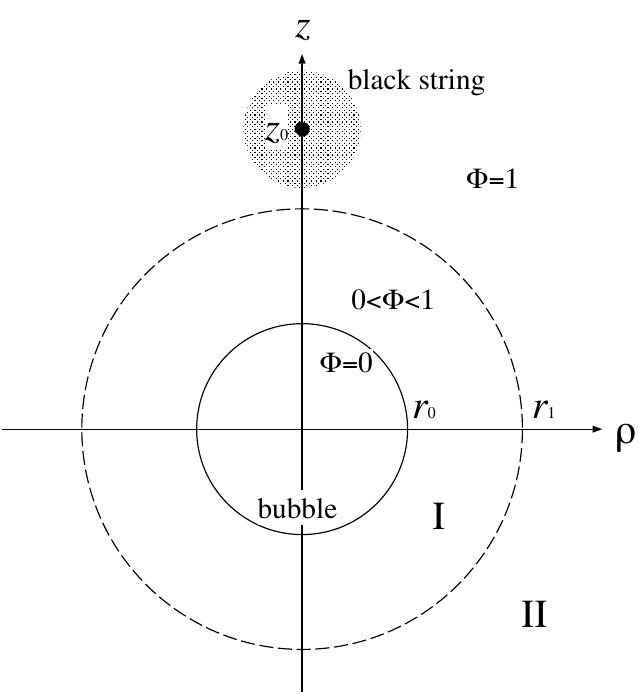} 
 \caption{The setup for the initial data for a black string and a KK bubble located at different positions.}
 \label{Fig:Setup}
\end{figure}
%

We adopt the metric ansatz of Eq.~\eqref{metric-ansatz2}
and solve Eq.~\eqref{equation-ansatz2} for $\Phi$ and $\Psi$ 
so that the generated initial data include
a black string and a KK bubble.
Figure~\ref{Fig:Setup} shows the schematic configuration 
of the system. In the 3D space $(x,\, y,\, z)$, we introduce
the spherical polar coordinates $(r,\,\theta,\,\phi)$ in the standard way.
The coordinate $\rho$ is introduced in Fig.~\ref{Fig:Setup} by $\rho=\sqrt{x^2+y^2}$. 
The KK bubble is supposed to be located at $r=r_0$, where $\Phi=0$
is imposed. The system is supposed to be axisymmetric,
and the functions $\Phi(r,\theta)$ and $\Psi(r,\theta)$ are
defined only in the region $r\ge r_0$.
The region $r\ge r_0$ is divided into two parts;
The region I $(r_0\le r< r_1)$ and the region II $(r_1\le r)$. 
In the region II, we require 
\begin{equation}
\Phi \ = \ 1 \quad (r_1\le r).
\end{equation}
One of the reasons for this requirement is that 
if $\Phi$ is changing its value near the puncture of the black string,
solving for $\Psi$ becomes very difficult. 
In the region I, we assume the form
\begin{equation}
\Phi \ = \ 1-\exp\left[\frac{r-r_0}{r-r_1}S(\theta)\right] \quad (r_0\le r<r_1),
\end{equation}
where $S(\theta)$ is a function to be determined numerically.
The introduction of $S(\theta)$ is necessary to realize the regularity 
on the bubble as discussed later. Note that the function $\Phi(r,\theta)$
introduced here is differentiable infinite times on $r=r_1$.

We suppose that the function $\Psi$ has the form
\begin{equation}
\Psi \ = \ \Psi_0 +\Delta\Psi,
\end{equation}
where
\begin{equation}
\Psi_0 \ = \ 1+\frac{M/2}{\sqrt{x^2+y^2+(z-z_0)^2}}.
\end{equation}
By choosing this form, the puncture is located at $z=z_0$
on the $z$-axis. We require $z_0>r_1$ so that
the puncture is located in the region II (where $\Phi=1$). 
The function $\Delta\Psi(r,\theta)$ must be determined numerically. 
The flat-space Laplacian is
\begin{equation}
\nabla^2= \partial_r^2 +\frac{2}{r}\partial_r +\frac{1}{r^2}\left(\partial_\theta^2+\cot\theta\partial_\theta\right),
\end{equation}
in the spherical-polar coordinates, and
the equation for $\Delta\Psi$ becomes
\begin{equation}
\nabla^2(\Delta\Psi) \ = \ 0 \quad (r_1\le r)
\end{equation}
in the region II (since $\Phi=1$).  In the region I, the equation for $\Delta\Psi$ becomes
\begin{equation}
\nabla^2(\Delta\Psi)
+\frac{1}{4\Phi}
\left[\Psi\nabla^2\Phi+2\Psi_{,r}\Phi_{,r}+\frac{2}{r^2}\Psi_{,\theta}\Phi_{,\theta}\right] 
\ = \ 0,
\label{Eq:DeltaPsi}
\end{equation}
where
\begin{subequations}
\begin{eqnarray}
\Phi_{,r} &=& \exp\left[\frac{r-r_0}{r-r_1}S(\theta)\right]\frac{r_1-r_0}{(r-r_1)^2}S(\theta),
\\
\Phi_{,rr} &=& \exp\left[\frac{r-r_0}{r-r_1}S(\theta)\right]\frac{r_1-r_0}{(r-r_1)^3}S(\theta)
\left[-2+\frac{r_0-r_1}{r-r_1}S(\theta)\right],
\\
\Phi_{,\theta} &=& -\exp\left[\frac{r-r_0}{r-r_1}S(\theta)\right]\frac{r-r_0}{r-r_1}S^\prime(\theta),
\\
\Phi_{,\theta\theta} &=& -\exp\left[\frac{r-r_0}{r-r_1}S(\theta)\right]\frac{r-r_0}{r-r_1}
\left[S^{\prime\prime}(\theta)+\frac{r-r_0}{r-r_1}\left(S^{\prime}(\theta)\right)^2\right].
\end{eqnarray}
\end{subequations}

We now discuss the boundary conditions.
On the bubble surface $r=r_0$, we have to impose two regularity
conditions. The first one comes from Eq.~\eqref{Eq:DeltaPsi}. 
Since $\Phi = 0$ on $r=r_0$, the inside of the square brackets
of Eq.~\eqref{Eq:DeltaPsi} must be zero. 
Since $\Phi_{,r}=S/(r_1-r_0)$, $\Phi_{,rr}=S(2-S)/(r_1-r_0)^2$, and $\Phi_{,\theta}=\Phi_{,\theta\theta}=0$ on $r=r_0$, we have
\begin{equation}
\Psi_{0,r}+\Delta\Psi_{,r} \ = \ 
-\frac{2r_1-r_0S(\theta)}{2(r_1-r_0)r_0}\,(\Psi_0+\Delta\Psi)
\qquad (r=r_0).
\label{boundary-condition1}
\end{equation}
The second one comes from the requirement that
there must be no conical singularity on the bubble.
The induced metric of the section $\theta$, $\phi=\mathrm{constnat}$ is
\begin{equation}
ds^2 \ \approx \
\left(\Psi(r_0,\theta)\right)^4 dr^2
+\left(\Phi_{,r}(r_0,\theta)\right)^2(r-r_0)^2d\chi^2.
\end{equation}
in the neighborhood of the bubble.
Therefore, $\chi$ must be identified with the period $\Delta\chi$ satisfying
\begin{equation}
\left(\frac{\Delta\chi}{2\pi}\right)\frac{S(\theta)}{r_1-r_0}
\ = \ \left(\Psi(r_0,\theta)\right)^2.
\label{boundary-condition2}
\end{equation}
Here, $\Delta\chi$ must be constant for arbitrary $\theta$. This is the reason
why $S(\theta)$ must be introduced: Without this function, the constancy 
of $\Delta\chi$ is incompatible with the second boundary condition.

Next, we discuss the outer boundary condition of $\Delta\Psi$.
Since the computation domain is limited, we impose the
outer boundary condition for a sufficient large $r_{\rm out}$.
Since $\Delta\Psi$ must behave as
\begin{equation}
\Delta\Psi \ \approx \ \frac{\Delta M}{2r},
\label{distant-behavior}
\end{equation}
where $\Delta M$ must be determined after solving the equation,
we impose the Robin boundary condition
\begin{equation}
\Delta\Psi_{,r}(r_{\rm out},\theta)
\ = \ 
-\frac{\Delta\Psi(r_{\rm out},\theta)}{r_{\rm out}},
\end{equation}
which can be obtained by eliminating $\Delta M$ from Eq.~\eqref{distant-behavior}.
After generating the solution,
the value of $\Delta M$ can be evaluated as
\begin{equation}
\Delta M \ = \ r_{\rm out}\int_0^\pi \Delta\Psi(r_{\rm out},\theta)\sin\theta d\theta.
\end{equation}
The ADM mass is given by $M_{\rm ADM} \ = \ M + \Delta M$.

\subsection{Numerical method and results}

We now discuss the numerical method for solving the above problem. 
There are five parameters, $\Delta\chi$, $r_0$, $r_1$,
$r_{\rm out}$, $M$, and $z_0$. We introduce 
the compactification radius at spacelike infinity, $R_\infty=\Delta\chi/2\pi$, and
adopt $R_\infty$ as the unit of the length, $R_\infty=1$.
Then, we choose the parameters  $r_0$, $r_1$,
$r_{\rm out}$, $M$, and $z_0$ by hand,
which corresponds to determining the system. 
In the numerical calculation, we adopt the coordinate
$x$ determined by
\begin{equation}
r \ = \ r_0 a^{x} \quad \mathrm{with} \quad a:=\frac{r_1}{r_0},
\end{equation}
in order to effectively span the distant place.
The coordinate range of $x$ is $0\le x\le x_{\rm out}$
where $r_{\rm out} = r_0a^{x_{\rm out}}$.
The surfaces $r=r_0$ and $r=r_1$ correspond to $x=0$
and $x=1$, respectively.

The equation for $\Delta\Psi$ becomes
\begin{multline}
\Delta\Psi_{,xx}+(\log a)\Delta\Psi_{,x}+(\log a)^2\left(\Delta\Psi_{,\theta\theta}
+\cot\theta \Delta\Psi_{,\theta}\right)
\\
+\frac{r^2(\log a)^2}{4\Phi}\left(\Psi\nabla^2\Phi+2\nabla\Psi\cdot\nabla\Phi\right) \ = \ 0.
\label{equation-DeltaPsi-x}
\end{multline}
Note that on the symmetry axis $\theta=0$ and $\pi$, we must use 
the regularized form of the equation that is derived using
\begin{equation}
\lim_{\theta\to0,\pi} \cot\theta \Delta\Psi_{,\theta}
\ = \ \Delta\Psi_{,\theta\theta}. 
\end{equation}
The first boundary condition of Eq.~\eqref{boundary-condition1} at the bubble surface $x=0$ 
is rewritten as
\begin{equation}
(\Delta\Psi)_{,x} \ = \ -\Psi_{0,x}
-\left(\frac{\log a}{2}\right)\,\frac{2r_1-r_0S(\theta)}{r_1-r_0}\left(\Psi_0+\Delta\Psi\right),
\label{boundary-condition1-x}
\end{equation}
and the second boundary condition of Eq.~\eqref{boundary-condition2} 
can be used without modification. 
The boundary condition at $x=x_{\rm out}$ is 
\begin{equation}
(\Delta\Psi)_{,x} \ = \ -(\log a)\left(\Delta\Psi\right).
\label{Robin-boundary-x}
\end{equation}

In order to solve for $\Delta\Psi(x,\theta)$ and $S(\theta)$,
the finite differencing method is used
by putting grid points at $x=I\Delta x$ with $I=0,1,...,I_{\rm max}$ 
and $\theta=J\Delta \theta$ with $J=0,1,..., J_{\rm max}$ 
in the coordinate domain $0\le x\le x_{\rm out}$
and $0\le \theta\le \pi$.
We mainly use the sixth-order finite difference equations,
but near the boundary, the accuracy is decreased to fourth-order.
The quantities $\Delta\Psi(x,\theta)$ and $S(\theta)$ are solved 
using the method of iteration: We first give a trial solution,
and make them gradually converge to a real solution. 
As the iteration method,
we used the successive-over-relaxation (SOR) method.
At each step of the iteration,
$S(\theta)$ is determined using the second boundary condition
on the bubble, Eq.~\eqref{boundary-condition2}.
With this value of $S(\theta)$, the value of $\Delta\Psi(x,\theta)$
of the next step is determined using Eq.~\eqref{equation-DeltaPsi-x}
with the boundary conditions of Eq.~\eqref{boundary-condition1-x}
and \eqref{Robin-boundary-x}.
As a result, $\Delta\Psi(x,\theta)$ and $S(\theta)$ simultaneously
converge to a real solution. 

%
\begin{figure}
 \centering
 \includegraphics[width=0.5\textwidth]{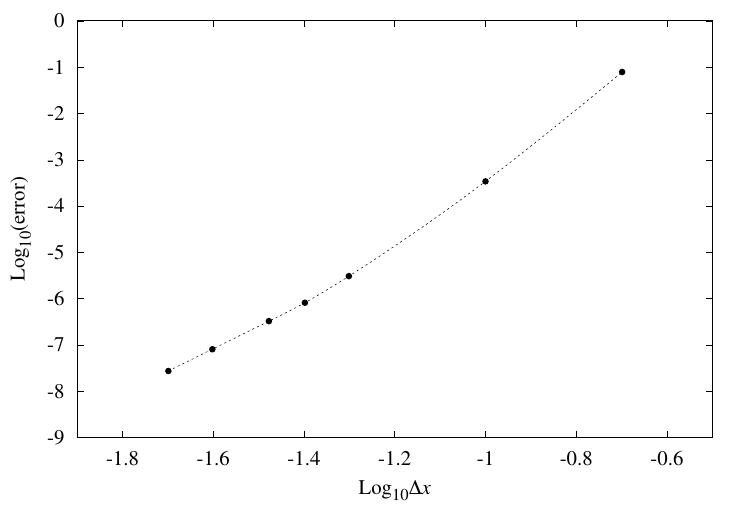} 
 \caption{The convergence of the numerical solution of $\Delta\Psi$ with respect to
 the grid size $\Delta x$.}
 \label{Fig:Convergence}
\end{figure}
%

%
\begin{figure}
 \centering
 \includegraphics[width=0.4\textwidth]{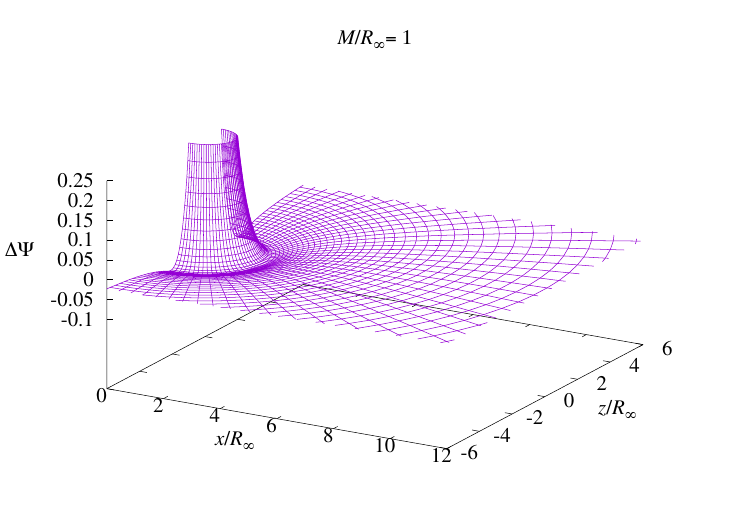} 
 \includegraphics[width=0.4\textwidth]{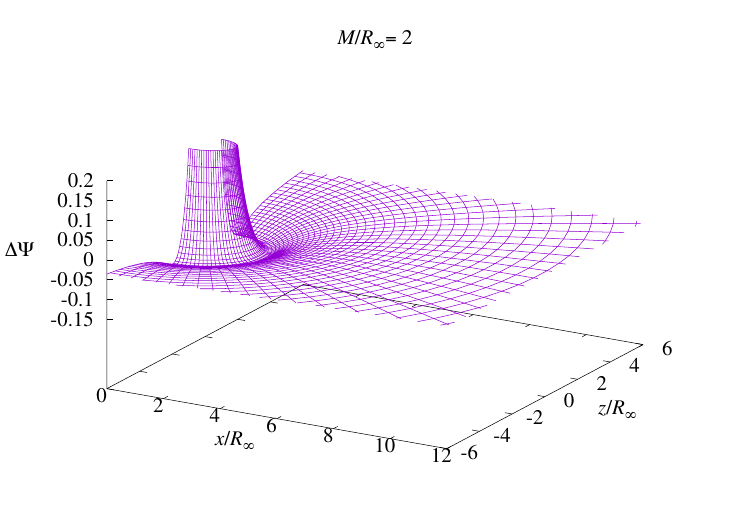} 
 \includegraphics[width=0.4\textwidth]{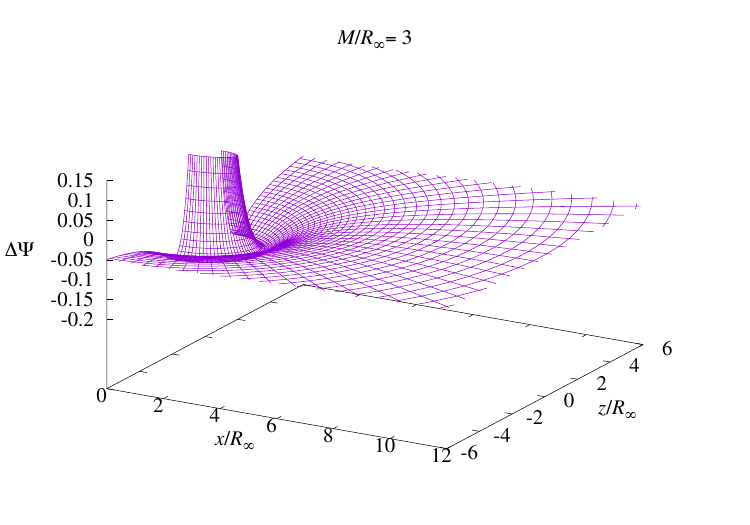} 
 \includegraphics[width=0.4\textwidth]{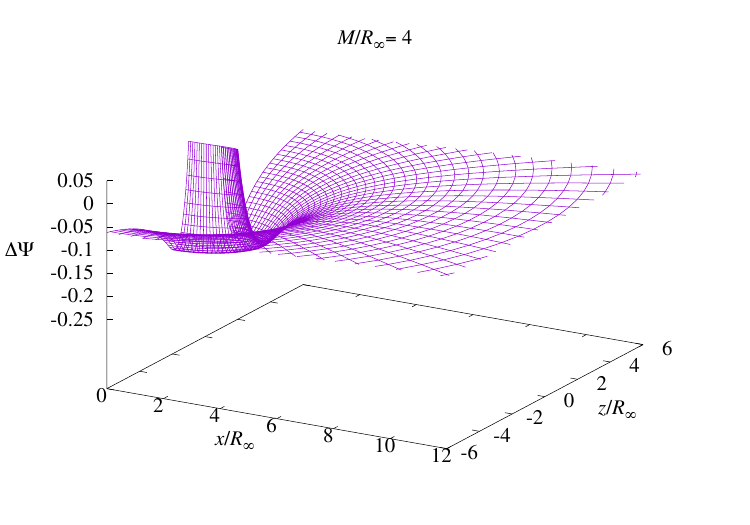} 
 \includegraphics[width=0.4\textwidth]{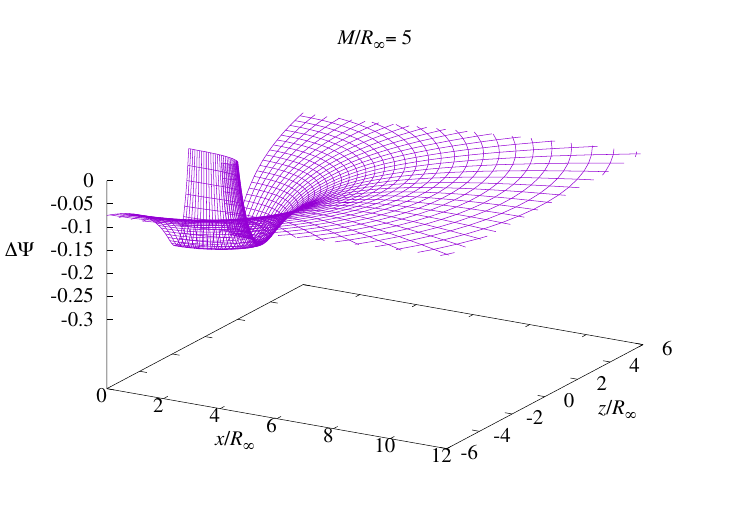} 
 \caption{3D plots of $\Delta\Psi$ as a function on the $(x,z)$-plane 
 for $M/R_\infty = 1$, $2$, $3$, $4$, and $5$. The other parameters are 
 $r_0/R_\infty = 1$, $r_1/R_\infty=3$, $z_0/R_\infty=5$
and $x_{\rm out}=10$.}
 \label{Fig:3D-plots}
\end{figure}
%

Before presenting the numerical results, 
we show the data that support the accuracy of our calculation. 
Figure~\ref{Fig:Convergence} shows the convergence of the numerical
solution of $\Delta\Psi$ with respect to the grid size $\Delta x$
for $r_0/R_\infty = M/R_\infty = 1$, $r_1/R_\infty=3$, $z_0/R_\infty=5$
and $x_{\rm out}=10$.
Here, the data taken with the grid numbers
$(I_{\rm max}, J_{\rm max})=(1000,200)$ (i.e. $\Delta x=0.01$) are adopted as the fiducial data
$\Delta\Psi^{\rm (f)}$,  
and the differences of the data for $(I_{\rm max}, J_{\rm max})=(500,100)$,
$(400,80)$, $(300,60)$, $(250,50)$, $(200,40)$, $(100,20)$, and $(50,10)$
(that is, $\Delta x$ ranges from $0.02$ to $0.2$)
are evaluated as
\begin{equation}
\mathrm{(error)} \ = \ \frac{\sum_{I,J}|\Delta\Psi-\Delta\Psi^{\rm (f)}|}{\sum_{I,J}|\Delta\Psi^{\rm (f)}|}.
\end{equation}
Here, the sum is taken over the grids that are located at the same points in the two calculations.
The figure shows the good convergence with respect to the grid size.

%
\begin{figure}
 \centering
 \includegraphics[width=0.5\textwidth]{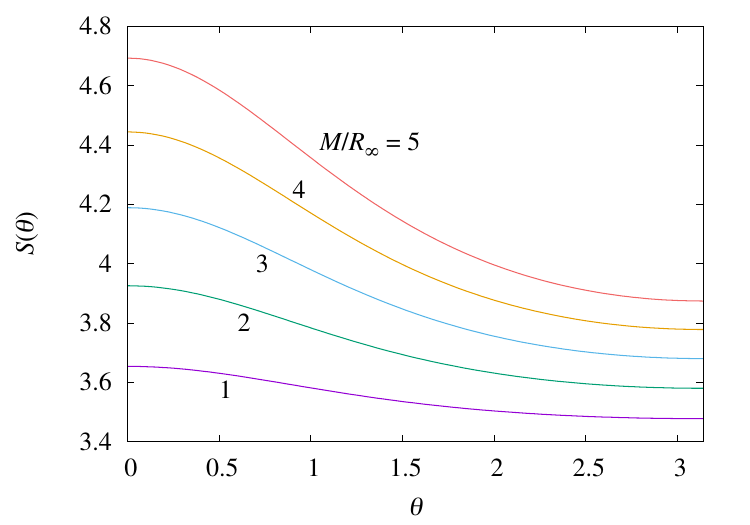} 
  \caption{The plot of $S(\theta)$ for the same parameters as Fig.~\ref{Fig:3D-plots}.}
 \label{Fig:S}
\end{figure}
%

%
\begin{figure}
 \centering
 \includegraphics[width=0.6\textwidth]{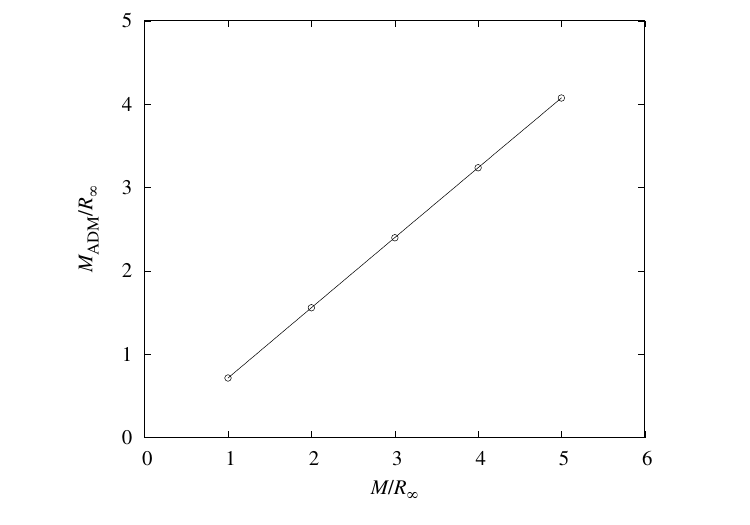} 
  \caption{The plot of $M_{\rm ADM}/R_\infty$ as a function of $M/R_\infty$
  for the same parameters as Fig.~\ref{Fig:3D-plots}.}
 \label{Fig:MADM}
\end{figure}
%

We now show the numerical results. 
Choosing $(I_{\rm max}, J_{\rm max})=(200, 45)$, 
we have generated initial data for
$M/R_\infty = 1$, $2$, $3$, $4$, and $5$ for
$r_0/R_\infty = 1$, $r_1/R_\infty=3$, $z_0/R_\infty=5$
and $x_{\rm out}=10$.
Figure~\ref{Fig:3D-plots} shows
the 3D plots of $\Delta\Psi$ on the $(x,z)$-plane. 
$\Delta\Psi$ takes negative values at the distant place.
Near the bubble surface, the sign of $\Delta\Psi$ depends
on the value of $M/R_\infty$, while $\Delta\Psi_{,r}<0$ for all cases.
Figure~\ref{Fig:S} shows the function of $S(\theta)$. 
$S(\theta)$ takes the maximum value at $\theta=0$, 
and the value of $S(\theta)$ for a fixed $\theta$ increases
as $M$ is increased.
Since $\Delta \Psi$ is negative at the distant place,
it gives a negative value of $\Delta M$. Hence,
the total ADM mass $M_{\rm ADM}$ is smaller than $M$
as indicated in Fig.~\ref{Fig:MADM}.

%
\begin{figure}
 \centering
 \includegraphics[width=0.7\textwidth]{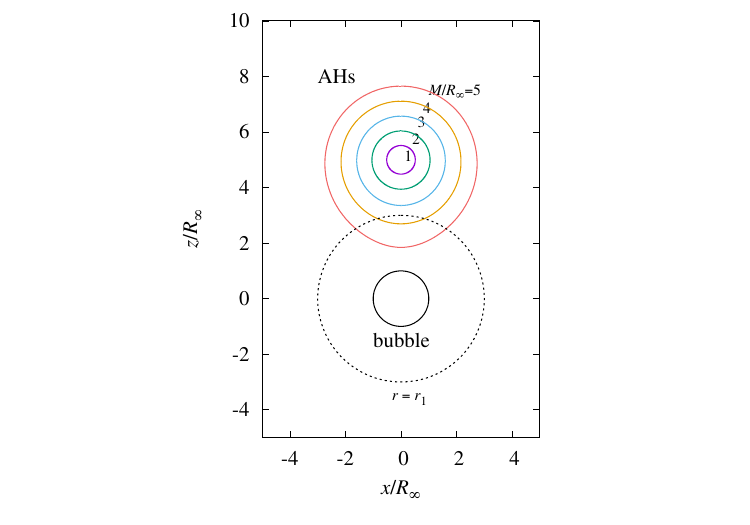} 
  \caption{The bubble surface and the apparent horizons of the black string
  on the $(x,z)$-plane 
  for the same parameters as Fig.~\ref{Fig:3D-plots}. Compare with Fig.~\ref{Fig:Setup}.}
 \label{Fig:bubble-AH}
\end{figure}
%

For each case, there is an apparent horizon that surrounds the puncture
at $z=z_0$ on the $z$-axis. By spanning a new spherical-polar coordinates
whose origin is located on the puncture,
the apparent horizon can be identified with the same method as 
the one discussed in Sec.~\ref{Sec:BL-2BS}.
Figure~\ref{Fig:bubble-AH} shows the shape of the bubble
and the apparent horizons of the black strings on the $(x,z)$-plane.
As $M$ is increased, the apparent horizon becomes larger, and
its coordinate shape becomes more distorted due to the presence of the bubble.

%
%
\section{Summary and discussion}
\label{Sec:Summary}

In this paper, we have studied the method of initial data construction
of KK spacetimes with an extra dimension whose radius is varying in space.
Specifically, we have presented the three types of the initial data.
The first one is the initial data with SO(3) symmetry
discussed in Sec.~\ref{Sec:Spherically-symmetric}; that is,
the initial data whose section $\chi=\mathrm{constant}$ is spherically symmetric. 
These initial data are divided into two subclasses:
the initial data without a KK bubble (Sec.~\ref{SubSec:without-bubble}) 
and those with a KK bubble (Sec.~\ref{SubSec:with-bubble}).
For the initial data without a KK bubble, on the one hand, the ADM mass is positive
and the area of the apparent horizon, which always exists, is bounded from above by that
of the Schwarzschild string with the same ADM mass.
On the other hand, in the initial data with a KK bubble, the apparent horizon
can be both present and absent depending on the system parameters.
When the KK bubble is not hidden inside the apparent horizon,
the ADM mass can be negative. The area of 
the apparent horizon, when it exists, can be larger than that of the
Schwarzschild string. The time evolution of the initial data with a KK bubble 
would be interesting since it may result in the formation of a naked singularity.

The second initial data discussed in Sec.~\ref{Sec:Brill-Lindquist}
is the analytic initial data with multiple black strings with
an extra dimension whose size varies in space. 
This is a natural extension of the Brill-Lindquist initial data
in the four-dimensional general relativity \cite{Brill:1963}. 
In particular, we have studied the condition for the apparent horizon formation
in the system of two black strings with equal masses, and have found
that it strongly depends on the structure of the extra dimension.
This would indicate that the structure of the extra dimension
plays an important role in the physics of black objects.

The third initial data are the system with a black string and a KK bubble
located at different positions 
discussed in Sec.~\ref{Sec:BS-KKbubble}.
Since we could not find an analytic solution to this system,
we have developed
the numerical method for constructing the initial data. 
This method has been successfully applied to some cases.
Since the KK bubble is not trapped inside the apparent horizon, these initial data 
are expected to provide us with a good initial condition
for studying the collision of an expanding KK bubble and
a black string with numerical relativity. 

This work is a starting point for studying the nonlinear
dynamics of an extra dimension whose compactification radius
depends on spatial positions.
In the previous work of the present author and a collaborator \cite{Yoshino:2009,Shibata:2010},  
the Baumgarte-Shapiro-Shibata-Nakamura (BSSN) formalism
and the Cartoon method for higher-dimensional spacetimes
have been developed  with successful applications. 
If the KK bubble is absent, this formalism is expected to be applied
to the initial data developed in this paper.
However, in the presence of the KK bubble, we need a method for treating
the surface of the KK bubble. Although such a method has been developed
for systems with SO(3) symmetry \cite{Sarbach:2003,Sarbach:2004},
this would be a challenging issue for systems without SO(3) symmetry.
We hope that these problems can be tackled in near future.

%
%
\acknowledgments

The author thanks Masato Nozawa, Shinya Tomizawa, Tetsuya Shiromizu, 
Hisa-aki Shinkai for helpful comments.
H. Y. is in part supported by JSPS KAKENHI Grant Numbers JP22H01220 and
JP21H05189,
and is partly supported by MEXT Promotion of Distinctive Joint Research Center Program  JPMXP0723833165.

\appendix






\begin{thebibliography}{99}

\bibitem{Overduin:1997}
J.~M.~Overduin and P.~S.~Wesson,
Phys. Rept. \textbf{283}, 303-380 (1997)
[arXiv:gr-qc/9805018 [gr-qc]].


\bibitem{Becker:2006}
K.~Becker, M.~Becker and J.~H.~Schwarz,
Cambridge University Press, 2006,
ISBN 978-0-511-25486-4, 978-0-521-86069-7, 978-0-511-81608-6.

\bibitem{Nishiwaki:2011}
K.~Nishiwaki, K.~y.~Oda, N.~Okuda and R.~Watanabe,
Phys. Lett. B \textbf{707}, 506-511 (2012)
[arXiv:1108.1764 [hep-ph]].

\bibitem{Choudhury:2016}
D.~Choudhury and K.~Ghosh,
Phys. Lett. B \textbf{763}, 155-160 (2016)
[arXiv:1606.04084 [hep-ph]].


\bibitem{Hou:2015}
S.~Hou, B.~Harms and M.~Cavaglia,
JHEP \textbf{11}, 185 (2015)
[arXiv:1507.01632 [hep-ph]].


\bibitem{ATLAS:2015}
G.~Aad \textit{et al.} [ATLAS],
JHEP \textbf{03}, 026 (2016)
[arXiv:1512.02586 [hep-ex]].

\bibitem{Murata:2014}
J.~Murata and S.~Tanaka,
Class. Quant. Grav. \textbf{32}, no.3, 033001 (2015)
[arXiv:1408.3588 [hep-ex]].

\bibitem{Westphal:2020}
T.~Westphal, H.~Hepach, J.~Pfaff and M.~Aspelmeyer,
Nature \textbf{591}, no.7849, 225-228 (2021)
[arXiv:2009.09546 [gr-qc]].

\bibitem{Witten:1981}
E.~Witten,
Nucl. Phys. B \textbf{195}, 481-492 (1982).


\bibitem{Brill:1989}
D.~Brill and H.~Pfister,
Phys. Lett. B \textbf{228}, 359-362 (1989).


\bibitem{Brill:1991}
D.~Brill and G.~T.~Horowitz,
Phys. Lett. B \textbf{262}, 437-443 (1991).


\bibitem{Shinkai:2000}
H.~a.~Shinkai and T.~Shiromizu,
Phys. Rev. D \textbf{62}, 024010 (2000)
[arXiv:hep-th/0003066 [hep-th]].


\bibitem{Sarbach:2003}
O.~Sarbach and L.~Lehner,
Phys. Rev. D \textbf{69}, 021901 (2004)
[arXiv:hep-th/0308116 [hep-th]].


\bibitem{Sarbach:2004}
O.~Sarbach and L.~Lehner,
Phys. Rev. D \textbf{71}, 026002 (2005)
[arXiv:hep-th/0407265 [hep-th]].


\bibitem{Corley:1994}
S.~Corley and T.~Jacobson,
Phys. Rev. D \textbf{49}, R6261-R6263 (1994)
[arXiv:gr-qc/9403017 [gr-qc]].

\bibitem{Chamblin:1996}
A.~Chamblin and R.~Emparan,
Phys. Rev. D \textbf{55}, 754-765 (1997)
[arXiv:hep-th/9607236 [hep-th]].

\bibitem{Horowitz:2002}
G.~T.~Horowitz and K.~Maeda,
Class. Quant. Grav. \textbf{19}, 5543-5556 (2002)
[arXiv:hep-th/0207270 [hep-th]].


\bibitem{Emparan:2001}
R.~Emparan and H.~S.~Reall,
Phys. Rev. D \textbf{65}, 084025 (2002)
[arXiv:hep-th/0110258 [hep-th]].

\bibitem{Elvang:2002}
H.~Elvang and G.~T.~Horowitz,
Phys. Rev. D \textbf{67}, 044015 (2003)
[arXiv:hep-th/0210303 [hep-th]].


\bibitem{Iguchi:2007}
H.~Iguchi, T.~Mishima and S.~Tomizawa,
Phys. Rev. D \textbf{76}, 124019 (2007)
[erratum: Phys. Rev. D \textbf{78}, 109903 (2008)]
[arXiv:0705.2520 [hep-th]].


\bibitem{Tomizawa:2007}
S.~Tomizawa, H.~Iguchi and T.~Mishima,
Phys. Rev. D \textbf{78}, 084001 (2008)
[arXiv:hep-th/0702207 [hep-th]].


\bibitem{Kunz:2008}
J.~Kunz and S.~Yazadjiev,
Phys. Rev. D \textbf{79}, 024010 (2009)
[arXiv:0811.0730 [hep-th]].

\bibitem{Yazadjiev:2009-1}
S.~S.~Yazadjiev and P.~G.~Nedkova,
Phys. Rev. D \textbf{80}, 024005 (2009)
[arXiv:0904.3605 [hep-th]].

\bibitem{Yazadjiev:2009-2}
S.~S.~Yazadjiev and P.~G.~Nedkova,
JHEP \textbf{01}, 048 (2010)
[arXiv:0910.0938 [hep-th]].

\bibitem{Nedkova:2010}
P.~G.~Nedkova and S.~S.~Yazadjiev,
Phys. Rev. D \textbf{82}, 044010 (2010)
[arXiv:1005.5051 [hep-th]].


\bibitem{Kunz:2013}
J.~Kunz, P.~G.~Nedkova and C.~Stelea,
Nucl. Phys. B \textbf{874}, 773-791 (2013)
[arXiv:1304.7020 [gr-qc]].



\bibitem{Astorino:2022}
M.~Astorino, R.~Emparan and A.~Vigan\`o,
JHEP \textbf{07}, 007 (2022)
[arXiv:2204.09690 [hep-th]].


\bibitem{Suzuki:2023}
R.~Suzuki and S.~Tomizawa,
Phys. Rev. D \textbf{109}, no.12, L121503 (2024)
[arXiv:2311.11653 [hep-th]].


\bibitem{Tomizawa:2024}
S.~Tomizawa and R.~Suzuki,
Phys. Rev. D \textbf{109}, no.10, 104067 (2024)
[arXiv:2403.16723 [hep-th]].


\bibitem{Copsey:2006}
K.~Copsey,
JHEP \textbf{12}, 007 (2007)
[arXiv:hep-th/0610058 [hep-th]].


\bibitem{Copsey:2007}
K.~Copsey,
JHEP \textbf{10}, 095 (2007)
[arXiv:0706.3677 [hep-th]].



\bibitem{Brill:1963}
D.~R.~Brill and R.~W.~Lindquist,
Phys. Rev. \textbf{131}, 471-476 (1963).

\bibitem{Lehner:2010}
L.~Lehner and F.~Pretorius,
Phys. Rev. Lett. \textbf{105}, 101102 (2010)
[arXiv:1006.5960 [hep-th]].


\bibitem{Figueras:2022}
P.~Figueras, T.~Fran\c{c}a, C.~Gu and T.~Andrade,
Phys. Rev. D \textbf{107}, no.4, 044028 (2023)
[arXiv:2210.13501 [hep-th]].


\bibitem{Yoshino:2009}
H.~Yoshino and M.~Shibata,
Phys. Rev. D \textbf{80}, 084025 (2009)
[arXiv:0907.2760 [gr-qc]].


\bibitem{Shibata:2010}
M.~Shibata and H.~Yoshino,
Phys. Rev. D \textbf{81}, 104035 (2010)
[arXiv:1004.4970 [gr-qc]].
  
  
\end{thebibliography}
\end{document}